\documentclass[10pt,conference]{IEEEtran}
\IEEEoverridecommandlockouts
\usepackage{tikz}
\usetikzlibrary{shapes,arrows,positioning,arrows.meta,quantikz}
\usepackage{cite}
\usepackage{amsmath,amssymb,amsfonts}
\usepackage{graphicx}
\usepackage{textcomp}
\usepackage{xcolor}
\usepackage{neuralnetwork}
\usepackage{subcaption}
\usepackage{xcolor}

\definecolor{erblue}{HTML}{0082F0}
\definecolor{erorange}{HTML}{FF8C0A}
\definecolor{ergreen}{HTML}{0FC373}
\definecolor{erpurple}{HTML}{AF78D2}
\definecolor{eryellow}{HTML}{FAD22D}
\definecolor{erred}{HTML}{FF3232}

\def\BibTeX{{\rm B\kern-.05em{\sc i\kern-.025em b}\kern-.08em
    T\kern-.1667em\lower.7ex\hbox{E}\kern-.125emX}}
    
\begin{document}

\colorlet{pink}{red!40}
\colorlet{blue}{cyan!60}
\colorlet{green}{green!60}

\newcommand{\red}[1]{{\color{red}#1}}
\newcommand{\bb}[1]{{\color{erred}Bence: #1}}
\newcommand{\zz}[1]{{\color{purple}Cimbi: #1}}
\newcommand{\nd}[1]{{\color{erblue}Dani: #1}}
\newcommand{\tzs}[1]{{\color{erorange}TZs: #1}}
\newcommand{\zk}[1]{{\color{ergreen}Zs\'ofi: #1}}
\newcommand{\vp}[1]{{\color{brown}VPeti: #1}}

\bstctlcite{IEEEexample:BSTcontrol} 


\title{Hybrid Quantum-Classical Autoencoders for End-to-End Radio Communication}

\author{\IEEEauthorblockN{
Zsolt Tabi\IEEEauthorrefmark{1}, 
Bence Bakó\IEEEauthorrefmark{1}\IEEEauthorrefmark{2}, 
Dániel T. R. Nagy\IEEEauthorrefmark{1}\IEEEauthorrefmark{3}, 
Péter Vaderna\IEEEauthorrefmark{3}, 
Zsófia Kallus\IEEEauthorrefmark{3},\\
Péter Hága\IEEEauthorrefmark{3}, 
Zoltán Zimborás\IEEEauthorrefmark{1}\IEEEauthorrefmark{2}}

\IEEEauthorblockA{
\IEEEauthorrefmark{1}Eötvös Loránd University, Budapest, Hungary \\
Email: zsolttabi@ik.elte.hu}
\IEEEauthorblockA{
\IEEEauthorrefmark{2}Wigner Research Centre for Physics, Budapest, Hungary \\
Email: bako.bence@wigner.hu, zimboras.zoltan@wigner.hu}
\IEEEauthorblockA{
\IEEEauthorrefmark{3}Ericsson Research, Budapest, Hungary\\
Email: \{daniel.a.nagy, peter.vaderna, zsofia.kallus, peter.haga\}@ericsson.com}}

\maketitle

\begin{abstract}
Quantum neural networks are emerging as potential candidates to leverage noisy quantum processing units for applications.
Here we introduce hybrid quantum-classical autoencoders for end-to-end radio communication.
In the physical layer of classical wireless systems, we study the performance of simulated architectures for standard encoded radio signals over a noisy channel. 
We implement a hybrid model, where a quantum decoder in the receiver works with a classical encoder in the transmitter part. Besides learning a latent space representation of the input symbols with good robustness against signal degradation, a generalized data re-uploading scheme for the qubit-based circuits allows to meet inference-time constraints of the application.

\end{abstract}


\section{Introduction}


One of the most popular Quantum Machine Learning (QML) methods are Quantum Neural Networks (QNNs) \cite{Schuld2014, PhysRevResearch.1.033063}. These are special variational quantum circuits, designed as the quantum analogues of classical neural networks. QNNs can be optimized with gradient-based or gradient-free optimization algorithms forming hybrid quantum-classical training loops \cite{Wierichs2022generalparameter, PennylanePaper}. Various QNN architectures have been proposed such as quantum convolutional neural networks \cite{QCNNWei2022}, generative models \cite{LLoydQGAN}, long short-term memories \cite{2021APSQLSTM} and autoencoders \cite{PhysRevA.102.032412,Romero_2017,PhysRevLett.124.130502}. Beside the high activity in algorithmic research within QML, their novel benchmarking and requirement setting applications are also motivating a wide variety of works  \cite{goldmansachs}.

Although QML is still in a phase of basic research with many open questions, its early implementations in wireless communication systems spark both scientific curiosity and commercial interest \cite{saladreview}. 
However, high-performance, near real-time applications might impose a new set of requirements on these solutions.

Wireless communication has undergone tremendous evolution during the last decades. The increasing adoption of AI and ML methods 
is opening up new development possibilities in various parts of the radio stack.
In the design of the sixth generation (6G) wireless networks, AI and ML technologies are considered to be tightly integrated into the system and smart algorithms can be applied to all aspects of network operations and procedures \cite{ericsson6G}.
Considering the improvement of quantum computers it is envisioned that quantum algorithms and especially QML will play a significant role in future networks \cite{saladreview}.

The structure of this paper is as follows.
In Sec.~\ref{sec:pae}, we give a high-level overview of wireless communication systems together with an autoencoder solution used in the radio physical layer. 
Our novel hybrid classical-quantum autoencoder prototype is presented in Sec.~\ref{sec:qpae}. We discuss our result in Sec.~\ref{sec:discussion}. 
Finally, we conclude with an outlook in Sec.~\ref{sec:conclusion}.

\section{Autoencoder Architecture in End-to-end Communication}
\label{sec:pae}

\begin{figure}[htpb]
\centering
\subfloat[][]{\includegraphics[trim={0 0.9cm 0 0.9cm}, clip,width=0.22\textwidth]{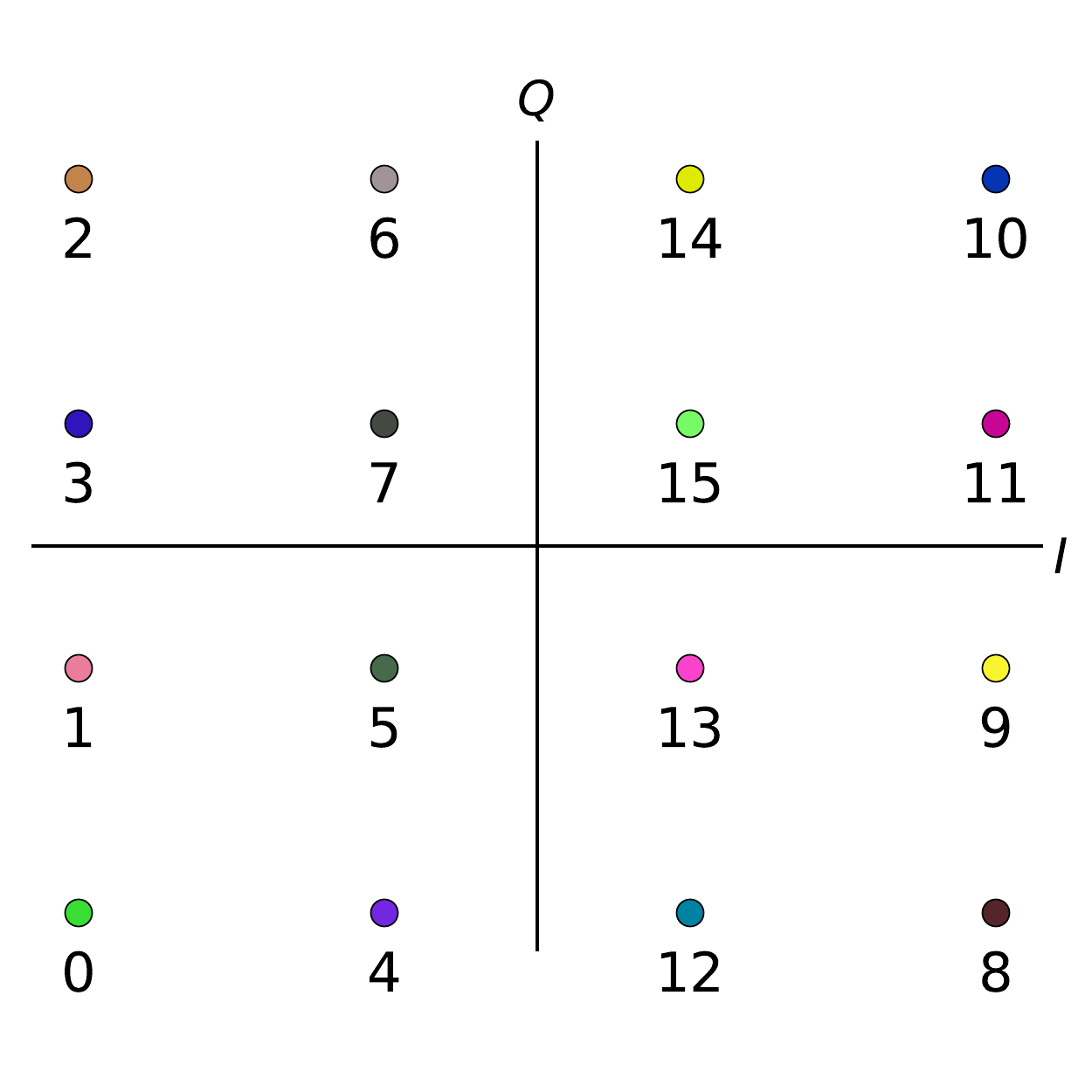}}%
\qquad
\subfloat[][]{\includegraphics[trim={0 0.9cm 0 0.75cm}, clip,width=0.22\textwidth]{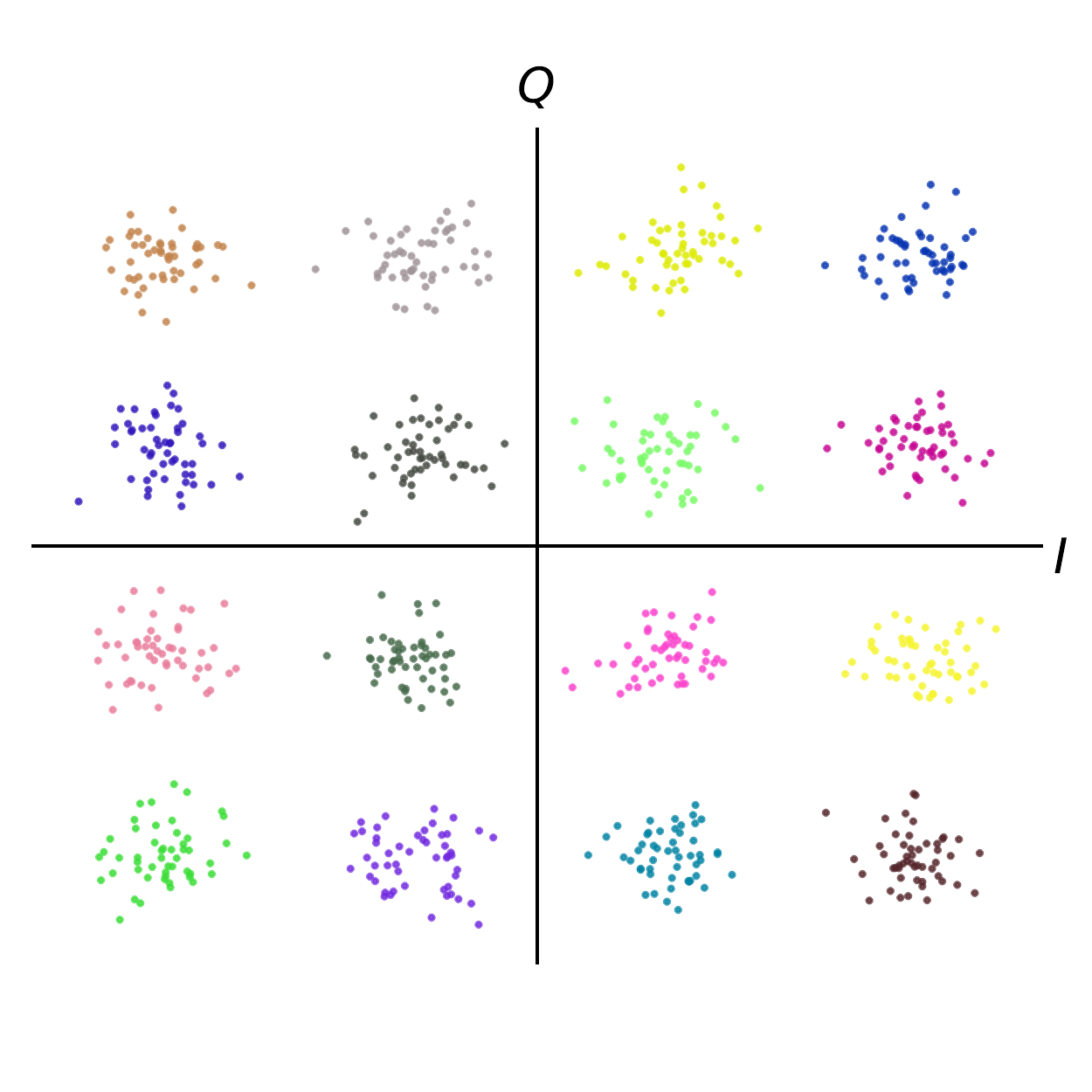}}
\caption{\textbf{16-QAM constellation diagrams.} (a) Transmitted symbols; (b) Received noisy signal.}%
\label{fig:const2}%
\end{figure}

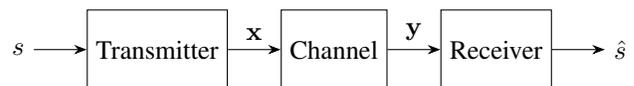
\begin{figure}[htpb]
\vspace{0.6cm}
\centering
\begin{tikzpicture}
    [block/.style={draw,minimum width=#1,minimum height=0em},
    block/.default=10em,high/.style={minimum height=3em},
    node distance=2em, > = Stealth]

    \node (n0) {${s}$};
    \node[block=3em,high,right=2em of n0] (n1) {Transmitter};
    \node[block=3em,high,right=of n1] (n2) {Channel};
    \node[block=3em,high,right=of n2] (n3) {Receiver};
    \node[right=2em of n3] (n4) {$\hat{s}$};

    \draw[->] (n0) -- (n1);
    \draw[->] (n1) -- node [above] {$\mathbf{x}$} (n2);
    \draw[->] (n2) -- node [above] {$\mathbf{y}$} (n3);
    \draw[->] (n3) -- (n4);
\end{tikzpicture}
\vspace{0.4cm}
\caption{\textbf{High-level representation of a communication system.} A message is transmitted through a noisy communication channel to be recovered by the receiver.} 
\label{fig:com_channel}
\end{figure}

\begin{figure*}[t!]
\centering
\includegraphics[height=0.2\textheight]{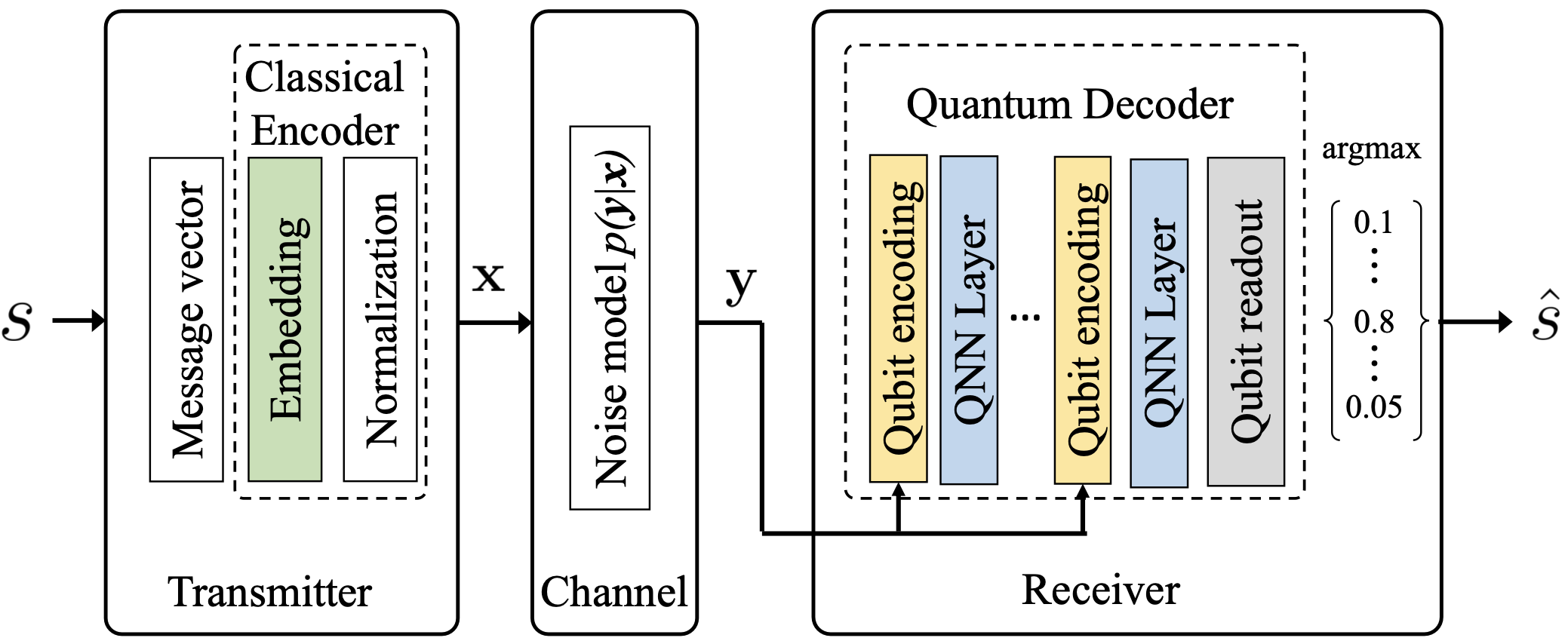}
\vspace{0.3cm}
\caption{\textbf{The proposed hybrid quantum-classical autoencoder embedded into the end-to-end communication architecture.} The transmitter maps each message $s$ to a symbol, then sends it through the channel. The channel is represented by an additive noise acting on the signal $\mathbf{x}$. The receiver, realized by a quantum decoder, consists of a multi-layer QNN adapting data re-uploading. It decodes the noisy signal and gives an estimate of the original message.}
\label{fig:architecture}
\end{figure*}

The components of communication networks are organized in a layered architecture where each layer is responsible for different communication aspects \cite{tanenbaum}.
The physical layer is the lowest layer. It provides the means of transmitting a stream of raw bits over a data channel connecting the network elements.
The physical layer on the transmitter side converts data in the form of bits to electromagnetic waves to be transmitted wirelessly, while the receiver converts the electromagnetic waves received by an antenna to binary data.
The main challenge in wireless data transmission is to overcome the channel impairments so that the messages can be recovered with small error rate.

The role of modulation is to convert digital data into radio waves.
It can be achieved in different ways, the information can be encoded by varying (shifting) either amplitude, 
frequency or phase of the electromagnetic wave. 
The more states the modulation has, the more bits are transferred in one symbol, resulting in higher data rate.
However, with higher order modulation the signal is more sensitive to channel errors, 
so the applied modulation usually depends on the channel quality.
Fig.~\ref{fig:const2} shows a constellation diagram of the symbol representation of 16-QAM modulation, where 4-bit strings can be represented as complex numbers in a scheme resistant to general noise patterns while achieving high data rate with minimal channel uses.

Based on \cite{8054694}, we model a simple communications system, shown in Fig.~\ref{fig:com_channel}, consisting of a \textit{transmitter} that modulates message $s \in \mathcal{M} = \{0,\ldots,M-1\}$ into a signal $\mathbf{x}$ and sends it over a noisy \textit{channel} to the \textit{receiver} that tries to decode the received signal $\mathbf{y}$ resulting in the received message $\hat{s}$. The transmitted signal $\mathbf{x}$ suffers degradation due to the noise present on the channel. 
In case of transmission over a complex channel with $n$ discrete channel uses, the transmitter can be represented as the transformation $f : \mathcal{M} \mapsto \mathbb{R}^{2n}$, mapping the message $s$ to $\mathbf{x} \in \mathbb{R}^{2n}$ signal with certain constraints imposed by the transmitting hardware (e.g., energy constraint or average power constraint). The channel can be modeled as a conditional probability density function $p(\mathbf{y}|\mathbf{x})$ that produces the output signal $\mathbf{y} \in \mathbb{R}^{2n}$ given the input signal, usually via some noise model (e.g., additive white Gaussian noise (AWGN)). 
The receiver is represented as the mapping $g : \mathbb{R}^{2n} \mapsto \mathcal{M}$ that recovers some estimate $\hat{s}$ of the original message from the received signal. 
In this work, we focus on the case of single channel use ($n=1$), however, this model can be easily adopted to cases of $n>1$. Also, the number of transmit-receive pairs can be increased to get a Multiple-Input and Multiple-Output (MIMO) system \cite{8054694, GARCIA2022192}.

Both of $f$ and $g$ transformations can be created in various ways. In case of simple noise models applied in the channel the transformations can be designed as explicit mathematical formulas. In the case of complex noise scenarios that are difficult to describe with mathematical models, a possible way is to train deep neural networks, especially autoencoders, to solve the encoding and decoding tasks \cite{8054694}. 

\begin{figure*}[t!]
    \centering
    \begin{subfigure}{0.45\textwidth}
        \centering
        \includegraphics[height=4cm]{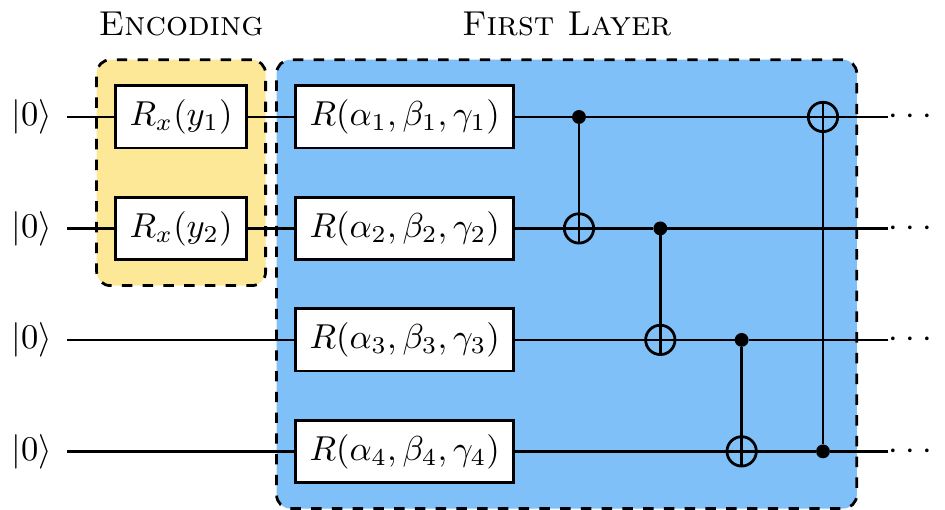}
        \caption{}
        \label{subfig:ans1}
    \end{subfigure}%
    ~ 
    \begin{subfigure}{0.45\textwidth}
        \centering
        \includegraphics[height=4cm]{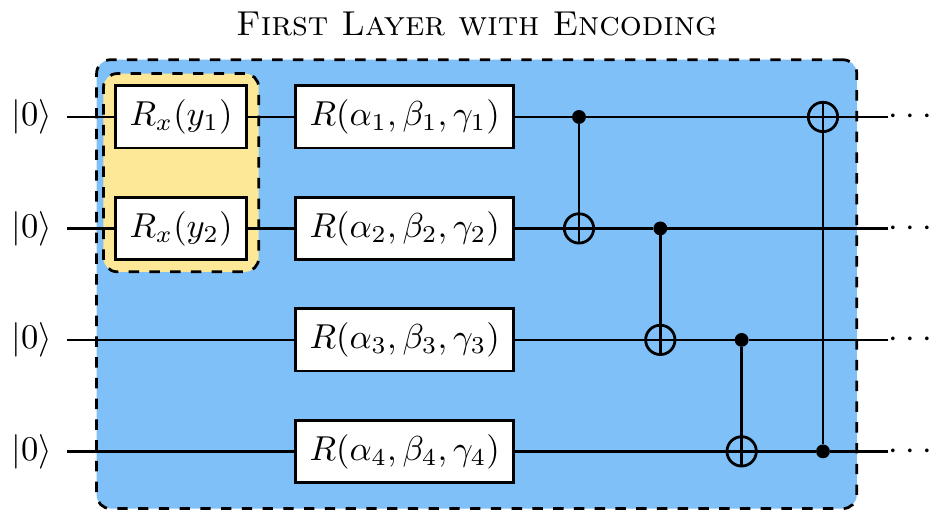}
        \caption{}
        \label{subfig:ans2}
    \end{subfigure}%
    \vspace{0.3cm}
    ~
    \begin{subfigure}{0.45\textwidth}
        \centering
        \includegraphics[height=4cm]{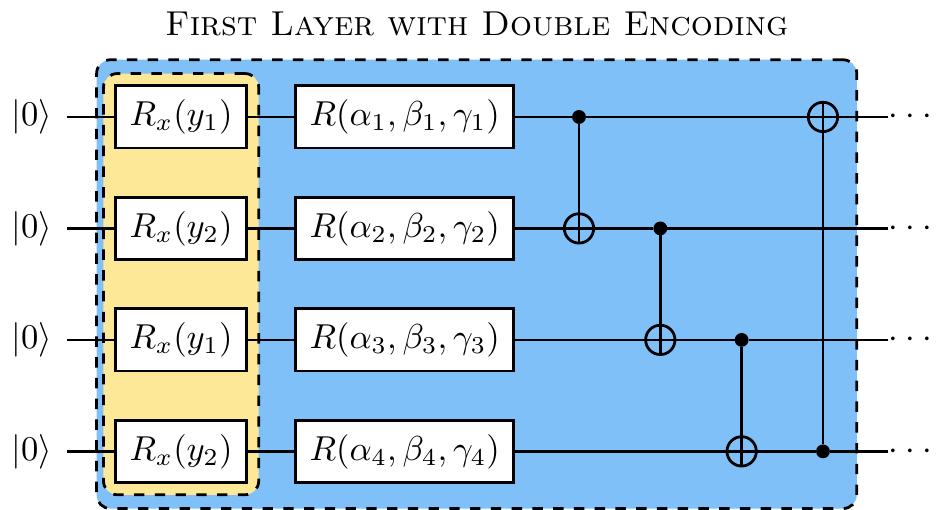}
        \caption{}
        \label{subfig:ans3}
    \end{subfigure}%
    ~
    \begin{subfigure}{0.45\textwidth}
        \centering
        \includegraphics[height=4cm]{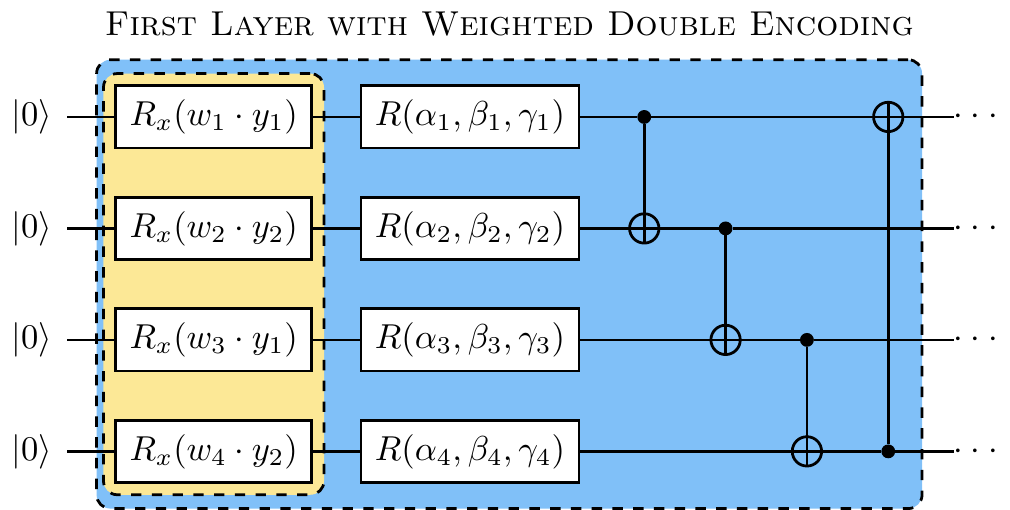}
        \caption{}
        \label{subfig:ans4}
    \end{subfigure}
    \caption{\textbf{Quantum decoder implementations with encoding schemes.}
    Ansatz circuits with (a) simple data encoding;
    (b) simple data re-uploading;
    (c) double data re-uploading;
    (d) weighted double data re-uploading.}
    \label{fig:ansatze}
\end{figure*}

An autoencoder is a special type of deep neural network, with the aim to compress or denoise data \cite{Hinton2006, Goodfellow-et-al-2016, NejiHala8893133}. Autoencoders consist of an encoder function $f : \mathbb R^D \mapsto \mathbb R^L$, and a decoder function 
$g : \mathbb R^L \mapsto \mathbb R^D$. 
The encoder transforms its input $\mathbf{\chi}$ into a \textit{latent space} representation $f(\mathbf \mathbf{\chi}) \in \mathbb R^L$, whereas the decoder tries to reconstruct it: $\hat{\mathbf{\chi}} = g(f(\mathbf{\chi}))$. 
Usually we have $L<D$, i.e., the encoder produces a compact representation of the data. 
$f$ and $g$ are typically deep neural networks trained jointly to minimize a loss function of the form $L(\mathbf{\chi}, g(f(\mathbf{\chi}))$. 

In telecommunication, opposite to the general compressing and denoising interpretation, autoencoders can be effectively used in the presented communication system to learn how to represent input messages as signals \cite{8054694}.  
This model differs from the ``typical'' autoencoder concept in the sense that it does not try to remove noise from the input, instead it learns how to represent the input in a way that is robust against a given noisy channel acting in the latent space of the autoencoder.
As a result of the training process, the latent space (or hidden layer) of the autoencoder contains the learned constellation of symbols (or codebook). The learned constellation is optimized for best mapping of the input messages to signals that can be accurately decoded with the largest success probability for the specific channel model. Whereas the encoder learns how to produce optimal symbols, the receiver learns how to decode these symbols after they have been corrupted by the channel, i.e., how to recover $\mathbf{x}$ after sampling from $p(\mathbf{y}|\mathbf{x})$.

\section{Hybrid Quantum Autoencoder for Radio Physical Layer}
\label{sec:qpae}

\subsection{Hybrid quantum autoencoder overview}
Quantum autoencoder architectures have previously been proposed to compress as well as denoise quantum data\cite{PhysRevLett.124.130502, PhysRevA.102.032412}.
Hybrid quantum-classical autoencoders enable many variations for quantum or classical encoding/decoding or the use of classical data.
In this work, a hybrid quantum-classical autoencoder is applied for processing classical information.
Building on the physical layer autoencoder presented in Sec.~\ref{sec:pae}, we propose a hybrid quantum-classical autoencoder with classical encoder on the transmitter side and a quantum decoder on the receiver side -- trained in an end-to-end solution. 
The encoder projects the original message to a lower dimensional representation, robust to the channel degradation effect. Once the signal is passed to the quantum decoder, the compressed information is mapped to a higher dimensional Hilbert space of the qubits by a QNN that has been previously shown to be efficient for classification tasks \cite{Farhi2018ClassificationWQ, Mari2020transferlearningin}.

In our model, the classical encoder consists of an embedding followed by a normalization.
A simple linear embedding is used to produce the constellation, satisfying the average power constraint by normalization.
The decoder is realized by a general strongly connected quantum neural network which we refer to as a \emph{quantum decoder}. 
By simulating increasing levels of noise in the channel, we can present a performance evaluation of the various neural network architectures.

\subsection{Quantum decoder architectures}\label{sec:qdec-arch}

A general QNN architecture has three main components as shown in Fig.~\ref{fig:architecture}: \emph{qubit encoding} for embedding the input data, the parameterized \emph{QNN layers}, and the \emph{qubit readout} given as a probability distribution over the possible constellation symbols obtained from suitable measurements with high enough number of shots.
To encode the output of the channel, we choose \emph{angle embedding} with parameterized $R_x$ rotations \cite{schuld2021machine}. With this embedding, there are multiple ways to encode two-dimensional feature vectors into four qubits.
As for the variational ansatz, we use strongly entangling layers introduced in quantum classifiers as they are known to be expressive reaching ‘wide corners of the Hilbert space’~\cite{PhysRevA.101.032308}. 
The measurements are performed in the computational basis and the obtained probability distribution over the $16$ basis states is the output of the decoder. 

The simplest single-layer realization of such a QNN structure is  presented in Fig.~\ref{subfig:ans1}.
To improve this ansatz, we can apply the data re-uploading trick recently introduced in \cite{PerezSalinas2020DataRF}. This technique, as shown in Fig.~\ref{subfig:ans2}, repeats the input encoding block before each layer of the QNN circuit. The intuition behind the effectiveness of this method is that by re-introducing the input before each layer, one can mimic the computational structure of typical classical deep neural networks, where the copying of the classical information is readily available, which would be, without this trick, prohibited by the no-cloning theorem in quantum machine learning. 
The expressivity of a model can be further increased by applying the encoding on different subsystems in parallel \cite{PhysRevA.103.032430}. With this in mind, we further enhance the ansatz by encoding the first feature into both qubit no.~1 and no.~3 and the second input feature into both qubit no.~2 and no.~4. 
This double data re-uploading ansatz is presented in Fig.~\ref{subfig:ans3}. 
As a final improvement, we considered the role of the number of trainable parameters. As the expressive power of the ansatz is highly dependent on the number of trainable parameters, one should try to include as many parameters as possible. 
One way to increase the number of parameters while keeping the circuit as shallow as possible -- to respect the limited hardware capabilities and the inference time constraints of the application -- is to introduce trainable weights in the data re-uploading blocks, as shown in Fig.~\ref{subfig:ans4}. 
This modification keeps the depth constant. 

\subsection{Training and fine-tuning}
For our hybrid autoencoder to achieve low estimation errors, the training of the end-to-end system requires to be further improved via hyper-parameter tuning.

First, the training of the hybrid model is done on batches uniformly sampled from the set of messages $\{0,\ldots,15\}$. These are
sent as two dimensional encoded symbols through the AWGN channel with SNR of $15$ dB and i.i.d. noise. 

The accuracy of the model is measured by evaluating the Symbol Error Rate (SER), a key performance indicator commonly used in radio communication. 
The network weight updates are calculated with the sparse categorical cross-entropy of the distribution generated by the decoder and the ground truth symbols. 
This loss function is used to calculate gradients in a mini-batch gradient descent with batch size of $64$ and Adam optimizer~\cite{kingma2014adam}. 
We simulate the hybrid autoencoder using PennyLane \cite{Bergholm2018PennyLaneAD}, a quantum machine learning framework with its TensorFlow \cite{tensorflow2015-whitepaper} backend. 
Second, we evaluate the reached model accuracy at various hyper-parameter settings. The search is conducted by KerasTuner \cite{omalley2019kerastuner} after partitioning the space as the simulator compute times are prohibitive of a full grid search.

We start by first evaluating the \emph{learning rate} parameter set $\eta \in \{0.1, 0.01, 0.001\}$ using the simple ansatz presented on Fig.~\ref{subfig:ans1} with $L=8$ layers with $1000$-shot measurements and $1000$ training steps. Based on these results, the only viable value of $\eta = 0.1$ is set for the rest of this study.

\begin{figure}[t!]
    \centerline{\includegraphics[width=0.45\textwidth]{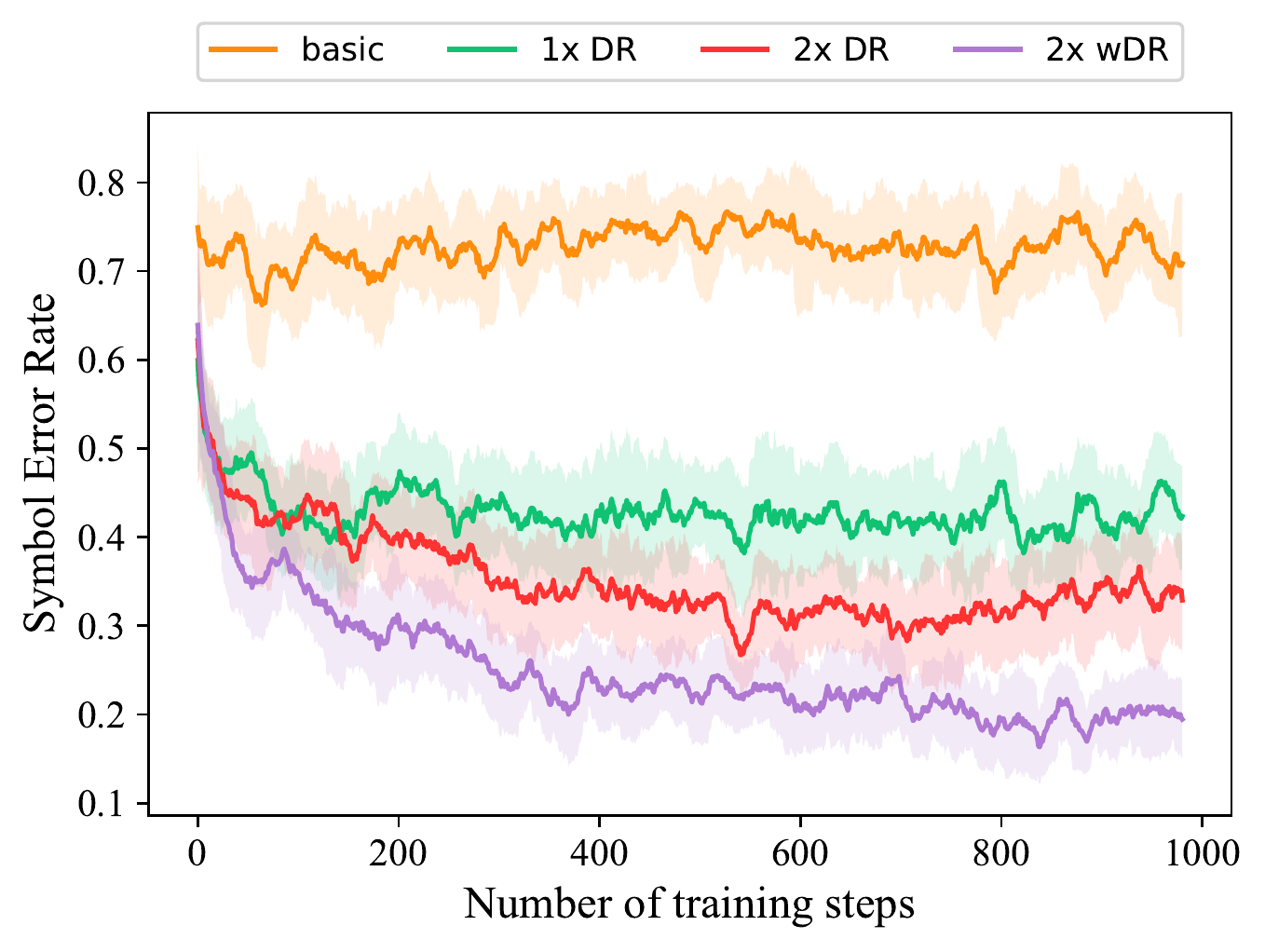}}
    \caption{\textbf{Learning curves of circuit architectures.} 
    The number of data re-uploading (none, single, double) and the weighted data encoding have high impact on the convergence properties of the quantum autoencoder.}
    \label{fig:dreup}
\end{figure}

We continue with evaluating modifications to the basic ansatz but keeping the number of layers $L=8$ and $1000$ training steps fixed, to minimize the overall computation time. The results are shown in Fig.~\ref{fig:dreup}. For the basic circuit, the SER fluctuates around its initial value without showing convergence to a desirable level.
A significant accuracy improvement of roughly $40\%$ is achieved by implementing single data re-uploading (Fig.~\ref{subfig:ans2} with ansatz of $1\times$DR). Introducing the double data re-uploading layer ($2\times$DR with ansatz of Fig.~\ref{subfig:ans3}) leads to another $15\%$ improvement. Finally, we can even further increase the performance by another $20\%$ when using the weighted double data re-uploading technique ($2\times$wDR with ansatz Fig.~\ref{subfig:ans4}).
Based on these results, the weighted double data re-uploading ansatz is chosen for further experiments. 

\begin{figure}[t!]
    \centerline{\includegraphics[width=0.45\textwidth]{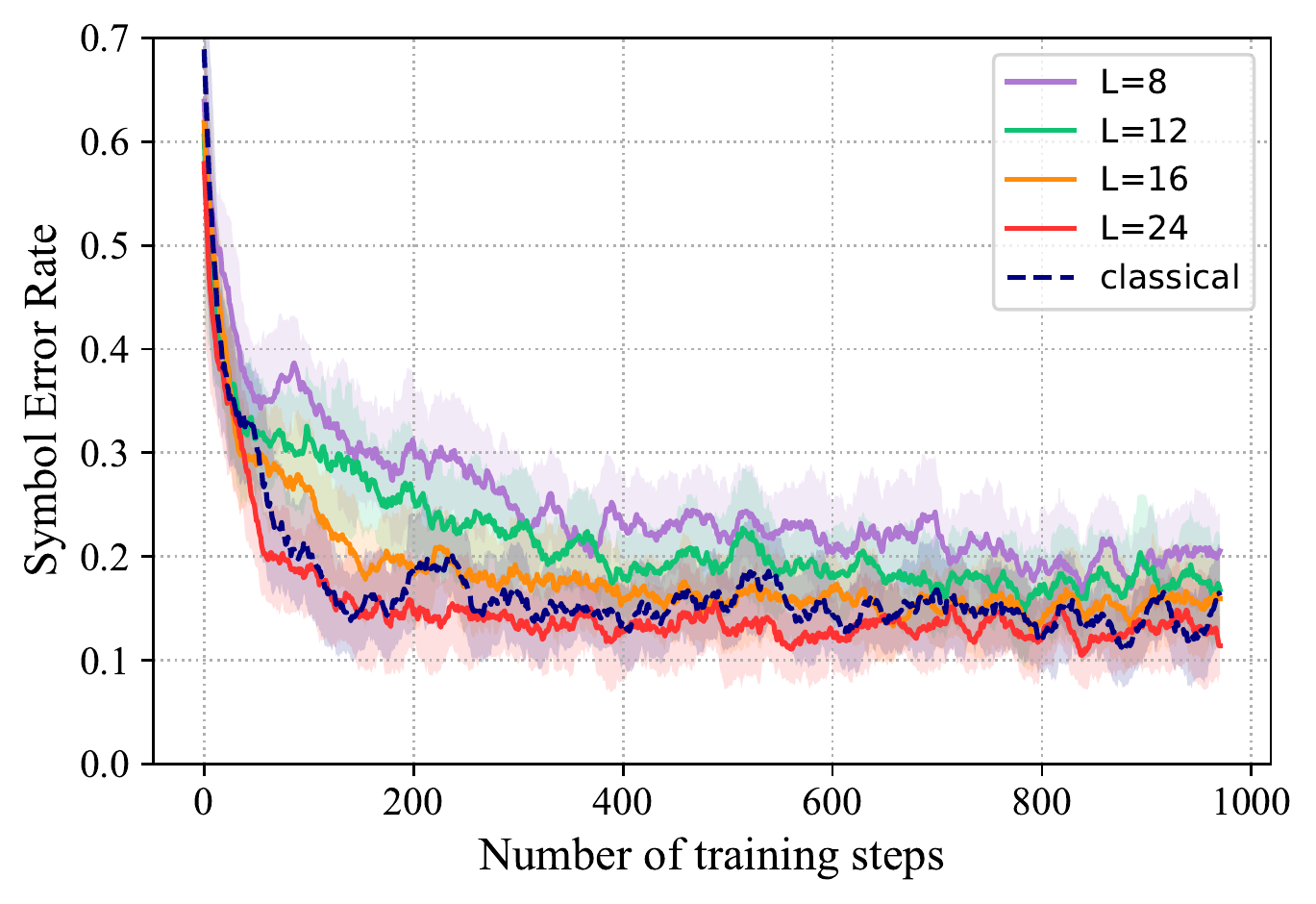}}
    \caption{\textbf{Learning curves of classical and hybrid autoencoders for a set of layer numbers.} We find that that the minimal number of layers necessary to achieve results comparable to the classical baseline is $16$. Throughout these tests, we used ansatz according to Fig.~\ref{subfig:ans4}. }
    \label{fig:learning}
\end{figure}

As a last step, we optimize the number of layers. The hybrid autoencoder using the best performing ansatz is trained with $8$ to $24$ layers. Increasing the number of layers clearly shows the improvement in SER as well as in convergence time as seen in Fig.~\ref{fig:learning}. 

\section{Performance Evaluation}
\label{sec:discussion}

\subsection{Validation}
Comparing our hybrid architecture to the classical method is crucial to validate the solution.
Based on the learning curves presented, the shallowest network reaching accuracy similar to the classical solution contains $L=16$ layers. 
Further increasing the number of layers leads to small improvements in accuracy but it is suboptimal in terms of circuit depth. 

\begin{figure}[t!]
    \centerline{\includegraphics[width=0.45\textwidth]{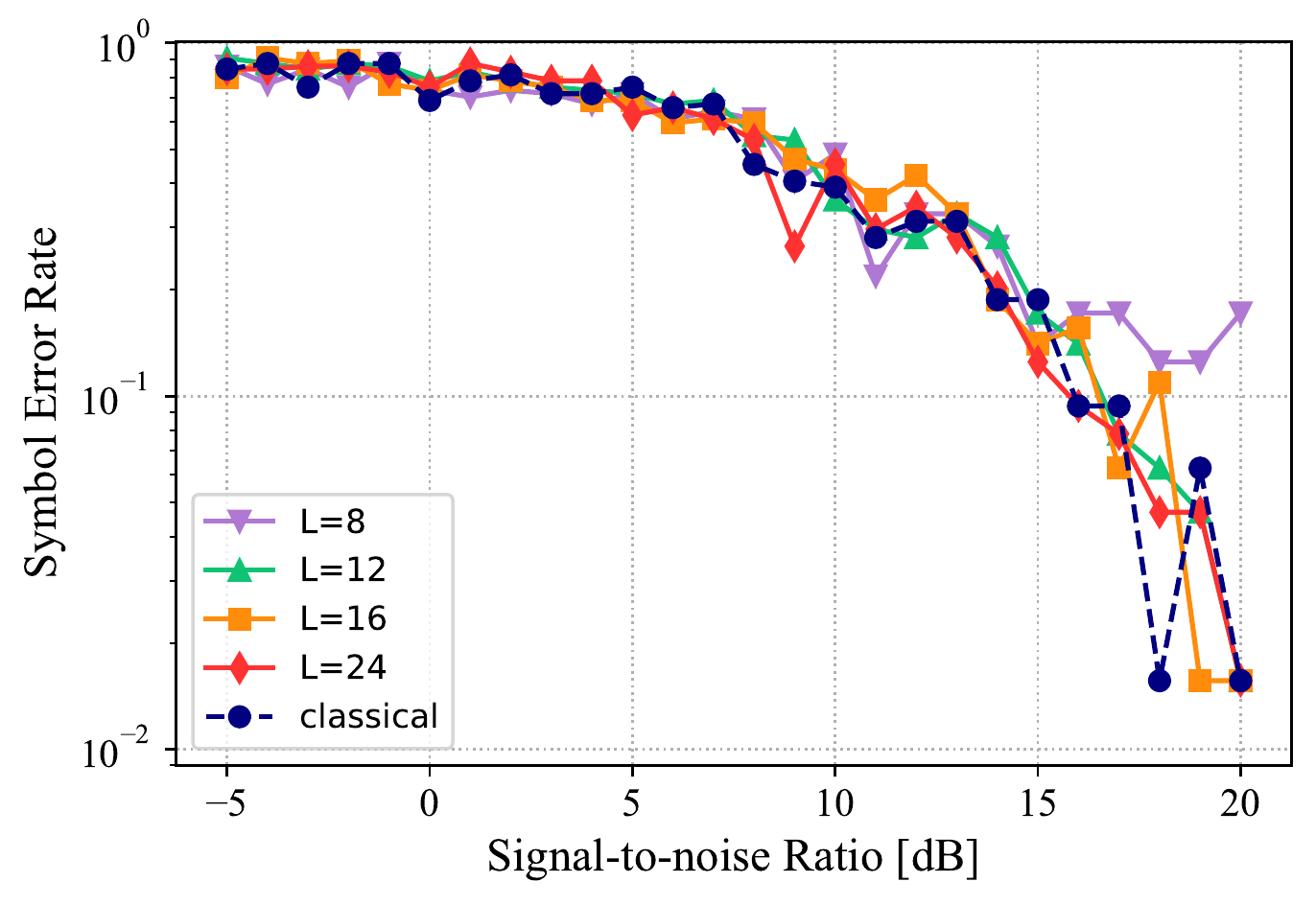}}
    \caption{\textbf{Validation of inference accuracy of the trained classical and hybrid autoencoders.} With increasing SNR values, hybrid models generalize to validation data on par with the classical.}
    \label{fig:minser}
\end{figure}

Although the hybrid quantum autoencoder models are trained at SNR of $15$ dB we further validate the results at different values. The evaluation is shown in Fig.~\ref{fig:minser}. 
We see that the trained networks generalize well on previously unseen SNR values, and reach performance similar to the classical baseline. 

\begin{figure}[t!]
    \centering
    \begin{subfigure}{0.25\textwidth}
    \centering
        \includegraphics[height=4cm]{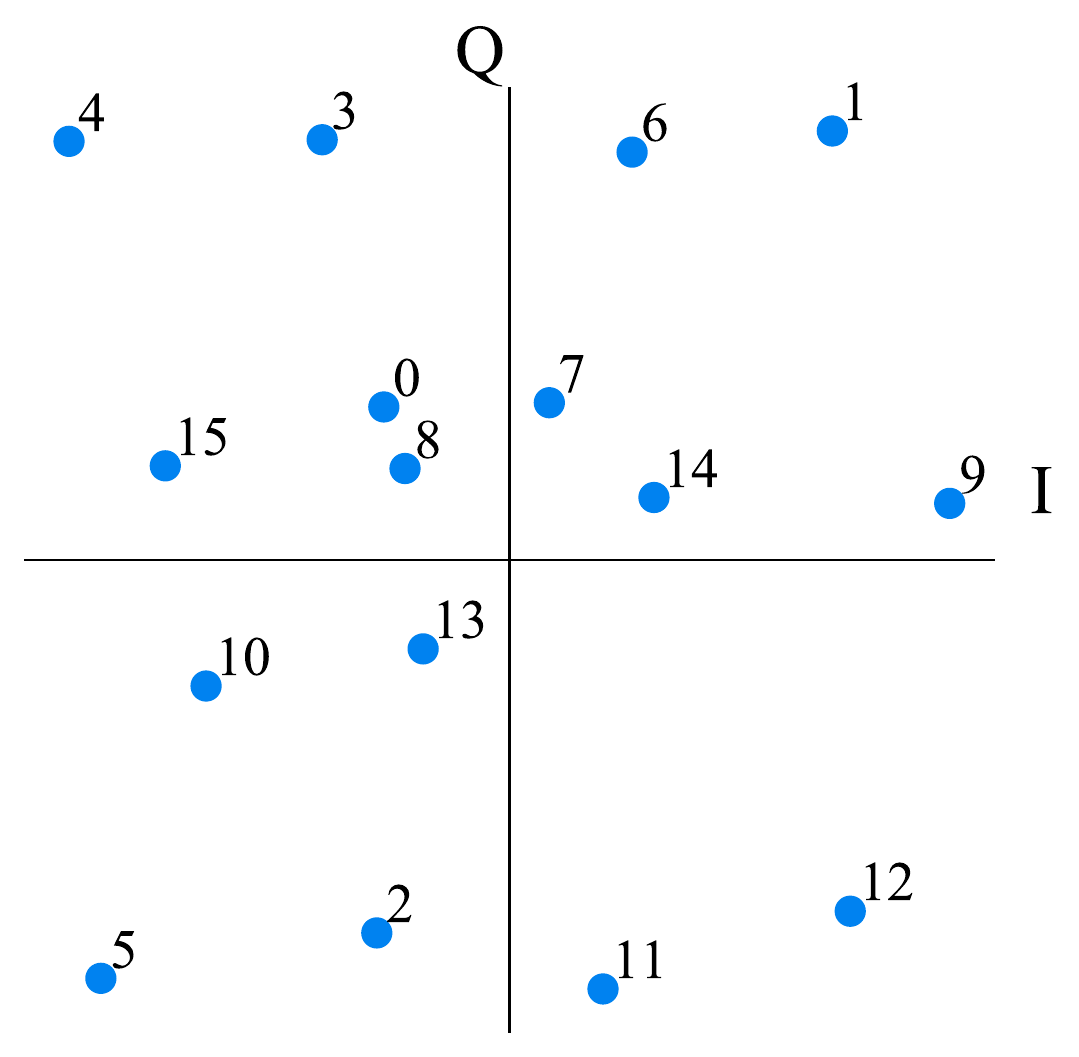}
        \caption{}
        \label{subfig:const3}
    \end{subfigure}%
    \begin{subfigure}{0.25\textwidth}
    \centering
        \includegraphics[height=4cm]{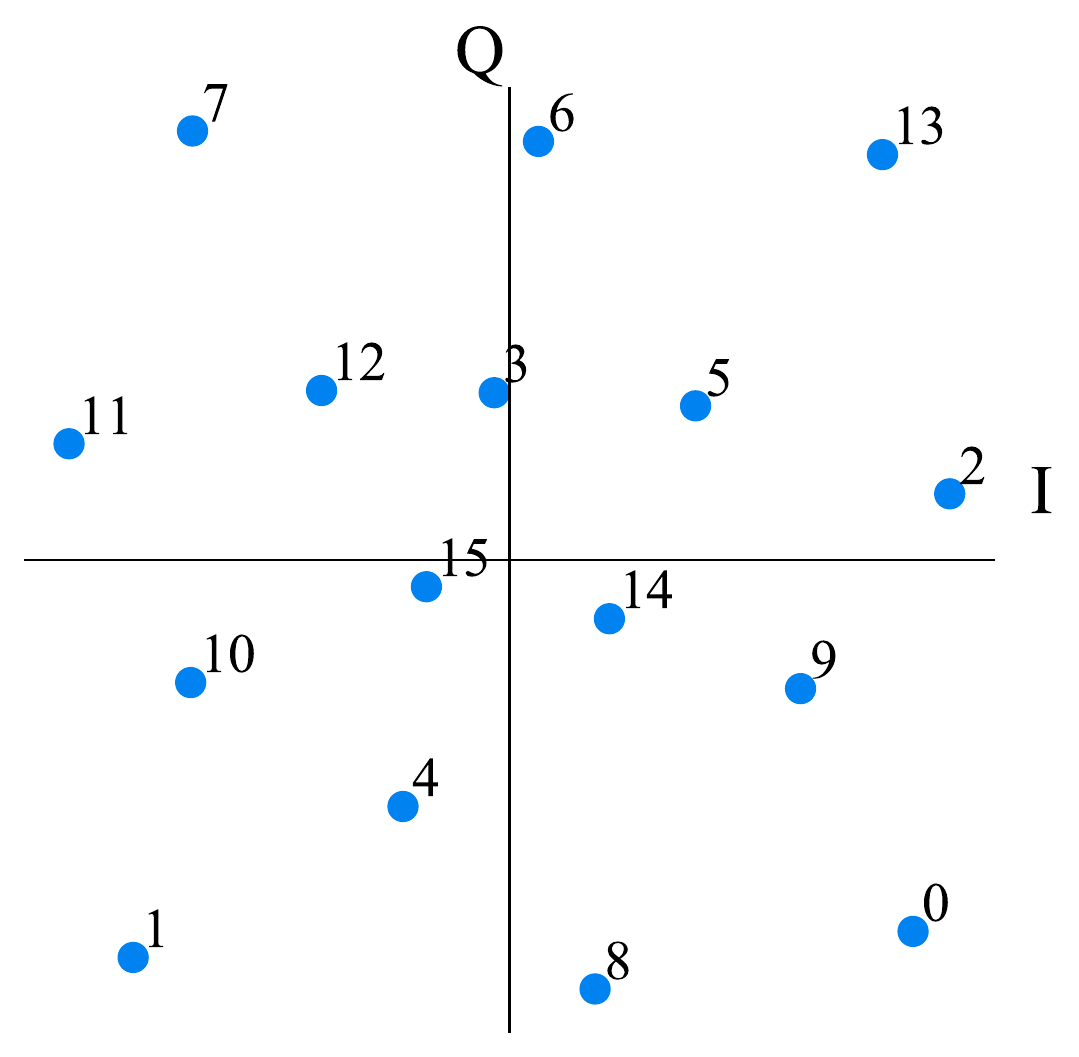}
        \caption{}
        \label{subfig:const4}
    \end{subfigure}%
    \caption{\textbf{Constellations (latent space representations) learned by the hybrid autoencoder trained with SNR=$15$.} (a) $L=8$ layers (b) $L=24$ layers. 
    \label{fig:const_result}
    }
\end{figure}

In Fig.~\ref{fig:const_result}, the constellation diagrams produced by autoencoders having different numbers of layers are shown. 
If the trained autoencoder has good performance, it is expected that the symbols are uniformly distributed in the diagram, similarly to Fig~\ref{fig:const2}.
We see that increasing the number of layers leads to a more balanced distribution of symbols in the $Q-I$ space, which implies that the symbols can be well separated in case of noisy channels.

\subsection{Time characteristics}
In radio telecommunication, the latency of the data transmission is also an important performance metric. In some use cases it is even critical that the end-to-end delay falls below a certain threshold. In 5G networks, it is possible to achieve ms level latency.
Hence, in addition to the accuracy it is inevitable to investigate the time characteristics of the autoencoder model.
After transpiling \cite{Qiskit} the circuit ansatz to IBM QPU backend \texttt{ibmq\_belem} and \texttt{ibmq\_santiago} \cite{ibmquantum} and constructing the pulse-level scheduling, we can calculate the theoretical execution times on both QPUs. The transpiled circuits are deeper than the original ansatz, because we need \texttt{SWAP} gates due to limited qubit connectivity and the basis gate-set of the device can differ from the one used in Fig~\ref{fig:ansatze}. In Table~\ref{tab:scaling}, we present the circuit depth and the approximate per shot execution times of quantum decoders depending on the number of layers. 
The time values in the table suggest the following feasibility considerations for running QNN in a real-time system. 
The number of shots highly determines the reliability of the result of the inference. When the quantum decoder is executed with 1000 shots (a level already acceptable in current systems for this problem size), the inference time is  the order of magnitude of $100ms$ which is higher than the accepted level in real-time radio systems. However, this can be reduced to the accepted level of below $10 ms$ because the probability distribution is expected to be highly peaked for well-trained autoencoders.

\section{Conclusion and Outlook}\label{sec:conclusion}

We presented a novel hybrid implementation of a quantum-classical autoencoder for end-to-end radio communication. The decoder was implemented as a variational quantum circuit.
We showed that the use of advanced double re-uploading encoding schemes allows for the inference-time constraints of the application to be met without losing accuracy required from the autoencoder.

By implementing a combination of parallel encodings and weighted data re-uploading, we showed how these schemes can improve not just the QNN expressivity but also the performance of the whole autoencoder model. 
We expect these quantum-enhanced models to outperform classical ones in more complex channel noise scenarios, a direction for future study.

\begin{table}[t!]
\renewcommand{\arraystretch}{1.3}
    \caption{\textbf{Estimated execution times of the quantum decoder.} The circuit was run on the \texttt{ibmq\_belem} and \texttt{ibmq\_santiago} depending on the number of layers, calculated with Qiskit's transpiler.}
    \label{tab:scaling}
    \centering
    \begin{tabular}{c|cc|cc}
    \hline
    \multicolumn{3}{c}{\hspace{1.5cm}\texttt{ibmq\_belem}} & \multicolumn{2}{c}{\texttt{ibmq\_santiago}} \\
    \hline\hline
        \# layers & depth & time [$\mu s/shot$] & depth & time [$\mu s/shot$] \\
        \hline
        8 & 125 & 54.3 & 145 & 30.4 \\
        12 & 187 & 78.4 & 221 & 43.6 \\
        16 & 260 & 111.8 & 297 & 56.9 \\
        20 & 311 & 124.2 & 373 & 70.12\\
        24 & 379 & 149.8 & 449 & 83.4 \\
    \hline \hline
    \end{tabular}
\end{table}

\section*{Acknowledgment}
Zsolt Tabi and Zimbor{\'a}s Zolt{\'a}n would like to thank the support of the Hungarian National Research, Development and Innovation Office (NKFIH) through the  Quantum Information National Laboratory of Hungary and through the Grants No. FK 135220, K124351 and TKP2021-NVA-29.

\bibliographystyle{IEEEtran}
\bibliography{IEEEabrv,final_refs}

\end{document}